\documentclass[twocolumn,floatfix,showpacs,pra,aps]{revtex4}

\usepackage{lastpage}
\usepackage{multirow}
\usepackage{graphicx,epsfig}
\usepackage{amsmath}
\usepackage{amssymb}
\usepackage{minitoc}  
\usepackage{amscd}
\usepackage{bm, rotating}
\usepackage{color}
\usepackage[normalem]{ulem}
\input{Qcircuit}

\DeclareMathSymbol{\Z}{\mathbin}{AMSb}{"5A}

\begin{document}

\title{Asymmetric quantum error correction via code conversion}

\author{Ashley M. Stephens$^{}\footnote{electronic address: a.stephens@physics.unimelb.edu.au}$, Zachary W. E. Evans, Simon J. Devitt, and Lloyd C. L. Hollenberg.}

\affiliation{
Centre for Quantum Computer Technology, School of Physics\\
University of Melbourne, Victoria 3010, Australia.}
\date{\today}

\begin{abstract}
In many physical systems it is expected that environmental decoherence will exhibit an asymmetry between dephasing and relaxation that may result in qubits experiencing discrete phase errors more frequently than discrete bit errors. In the presence of such an error asymmetry, an appropriately asymmetric quantum code - that is, a code that can correct more phase errors than bit errors - will be more efficient than a traditional, symmetric quantum code. Here we construct fault tolerant circuits to convert between an asymmetric subsystem code and a symmetric subsystem code. We show that, for a moderate error asymmetry, the failure rate of a logical circuit can be reduced by using a combined symmetric asymmetric system and that doing so does not preclude universality.
\end{abstract}


\maketitle
In any large quantum computer - a computer consisting of a large number of confined, controllable, and measurable two-level quantum systems - it is reasonable to assume that fault tolerant quantum error correction \cite{Shor2, Steane1} will be will be required to detect and correct errors induced by decoherence and systematic imprecision that cannot be suppressed by other means. The theory of fault tolerant quantum error correction is now well understood \cite{Aliferis4}, and in recent years there have been several important developments that have brought this theory closer to experimental reality. This has been achieved by considering, for example, systems where interactions are restricted to neighboring qubits \cite{Svore1} and systems where measurement is assumed to be relatively slow \cite{DiVincenzo1}.

In many systems it is expected that qubits will be affected by dephasing (loss of phase coherence) more strongly than by relaxation (exchange of energy with the environment) \cite{Vandersypen1, Astafiev1}. Quantum error correction causes dephasing to be manifested as discrete phase errors and relaxation as discrete bit errors. Traditional quantum error correction protocols are symmetric with respect to the phase and bit bases and so enable the detection and correction of an equal number of phase and bit errors during each cycle. As the protection afforded by quantum error correction is achieved at the expense of resources - time and qubits - this implies that some fraction of these resources is wasted in attempting to detect and correct errors that may be relatively unlikely to occur. This is made worse by the fact that the inclusion of unnecessary circuitry will actually increase the probability that an error will occur.

In light of this knowledge it is possible to increase the efficiency of traditional, symmetric quantum error correction by independently adjusting the frequencies of phase and bit error syndrome extraction to reflect a known error asymmetry \cite{Evans1}. A potentially more powerful tool is an asymmetric quantum code - that is, a code that is able to detect and correct more phase errors than bit errors during each cycle. It is known that such codes can be constructed \cite{Calderbank1, Ioffe1, Bacon2, Aly1, Sarvepalli1} but it is less obvious that they are suitable for universal fault tolerant quantum computation \cite{Gourlay1}. 

Here we detail the construction of fault tolerant circuits to convert between an asymmetric code and a symmetric code. With these circuits the asymmetric code can be used whenever possible and existing fault tolerant gate constructions for the symmetric code can be used to achieve universality. We show that, for a moderate error asymmetry at all locations, the failure rate of sections of a logical circuit can be reduced by up to two orders of magnitude by using a combined symmetric asymmetric system in this way. The shortcoming of our method is that it requires gates that are non-diagonal in the computational basis. Errors in these gates are not expected be strongly asymmetric, at least at the physical level. This issue has been addressed in a subsequent paper by Aliferis and Preskill in which circuits for asymmetric error correction are constructed entirely from gates that are diagonal in the computational basis \cite{Aliferis3}.

\begin{figure*}[ht]
\begin{center}
\resizebox{82mm}{!}{\includegraphics{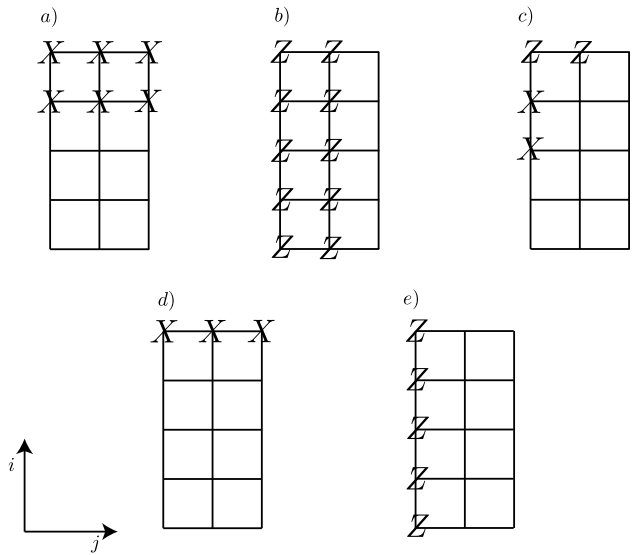}}
\end{center}
\vspace*{-10pt}
\caption{Examples of group elements for each relevant subsystem of the asymmetric subsystem code, $\mathcal{C}$($5$,$3$). Each vertex point represents one of the 15 physical qubits needed to encode a single logical qubit.  Figs. a) and b) illustrate $X$ and $Z$ elements from the stabilizer group $\mathcal{S}$. Fig. c) illustrates two commuting elements from the non-Abelian gauge group, $\mathcal{T}$.  Finally, Figs. d) and e) illustrate the two Pauli gates for the single logically encoded qubit from the group, $\mathcal{T}$.}
\label{figure:set}
\end{figure*}

Our starting point is the Bacon-Shor subsystem code, $\mathcal{C}$($n_1$,$n_2$), a stabilizer CSS quantum code that encodes one logical qubit into $n_1n_2$ physical qubits \cite{Bacon2, Shor1, Poulin1}. It is instructive to consider the qubits that make up the logical qubit as the vertices of an $n_1$$\times$$n_2$ grid. With this in mind, the group structure of the code is separated into three relevant subsystems. The first is the stabilizer group \cite{Gottesman1}, $\mathcal{S}$, which is generated by the operators
\begin{equation}
\begin{aligned}
\mathcal{S} = \langle X_{i,*} X_{i+1,*} ; Z_{*,j}Z_{*,j+1} \ | \ i \in \Z_{n_1-1} ; j \in \Z_{n_2-1} \rangle,
\end{aligned}
\label{stabilizers:bs}
\end{equation}
where we have retained the notation used in \cite{Aliferis1}, $Z$ and $X$ represent the Pauli matrices $\sigma_Z$ and $\sigma_X$ respectively, $U_{i,*}$ and $U_{*,j}$ represent an operator, $U$, acting on all qubits in a given row, $i$, or column, $j$, respectively, and $\Z_n=\{1,\dots,n\}$.  The second relevant subsystem is known as the gauge group, $\mathcal{T}$, and is described by the non-Abelian group generated by the pairwise operators
\begin{equation}
\begin{aligned}
\mathcal{T} =     &\langle X_{i,j}X_{i+1,j} \ | \ i \in \Z_{n_1-1} ; j \in \Z_{n_2} \rangle, \\
			&\langle Z_{i,j}Z_{i,j+1} \ | \ i \in \Z_{n_1} ; j \in \Z_{n_2-1} \rangle. 
\end{aligned}
\label{gauge}
\end{equation}
The third relevant subsystem is the logical space, $\mathcal{L}$, which can be defined via the logical Pauli operators
\begin{equation}
\mathcal{L} = \langle  Z_{*,1} ; X_{1,*} \rangle,
\end{equation}
which when combined with $\mathcal{S}$ form a non-Abelian group. Fig.~\ref{figure:set} illustrates elements of the asymmetric subsystem code, $\mathcal{C}$($5$,$3$), from each of the three groups, $\mathcal{S}$, $\mathcal{T}$, and $\mathcal{L}$.

The structure of the Bacon-Shor subsystem code arises from the group relations of these three subsystems.  If we let $\mathcal{H}$ denote the Hilbert space of the $n_1n_2$ physical qubits, $\mathcal{S}$ forms an Abelian group and hence can act as a stabilizer set denoting subspaces of $\mathcal{H}$.  If we describe each of these subspaces by a binary vector, $\vec{e}$, formed from the eigenvalues of the stabilizers, $\mathcal{S}$, then each subspace splits into a tensor product structure so that
\begin{equation}
\mathcal{H} = \bigoplus_{\vec{e}} \mathcal{H}_{\mathcal{T}} \otimes \mathcal{H}_{\mathcal{L}},
\end{equation}
where elements of $\mathcal{T}$ act only on the subsystem $\mathcal{H}_{\mathcal{T}}$ and the operators $\mathcal{L}$ act only on the subsystem $\mathcal{H}_{\mathcal{L}}$, in which the single logical qubit is stored. The operators in $\mathcal{L}$ are logical $X$ and $Z$ gates on this qubit. The stabilizers, $\mathcal{S}$, can be decomposed as
\begin{equation}
\begin{aligned}
&S_{X_i} = \bigotimes_{j=0}^{n_2} X_{i,j}X_{i+1,j} \ ; \ i \in \Z_{n_1-1},\\
&S_{Z_j} = \bigotimes_{i=0}^{n_1} Z_{i,j}Z_{i,j+1} \ ; \ j \in \Z_{n_2-1},
\end{aligned}
\label{stabilizers:bs2}
\end{equation}
that is, as products of elements of $\mathcal{T}$. Therefore, the eigenvalues of the stabilizers can be determined by measuring the eigenvalues of the gauge operators. This will potentially perturb the gauge state of the system but will not affect the information stored in $\mathcal{H}_{\mathcal{L}}$. Knowing these eigenvalues is sufficient to detect and correct at least $\lfloor\frac{n_1-1}{2}\rfloor$ $Z$ errors and at least $\lfloor\frac{n_2-1}{2}\rfloor$ $X$ errors in the encoded qubit, defining $\mathcal{C}(n_1,n_2)$ as having $Z$ and $X$ distances of $n_1$ and $n_2$ respectively. As each of the operators in Eq.~\ref{gauge} are pairwise, this simplifies the construction of fault-tolerant error correction circuits \cite{Bacon1, Aliferis1, Aliferis5}.

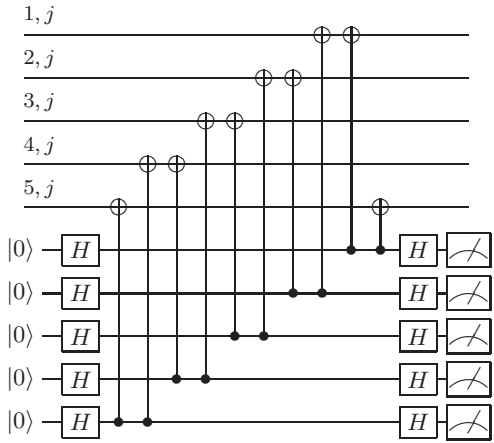
\begin{figure}
\[ \Qcircuit @C=0.4em @R=0.36em @!R {               
      								& \qw & \qw 		& \qw 	& \qw 	& \qw 	& \qw 	& \qw 	& \qw 	& \qw 	& \targ	& \targ	& \qw	& \qw 	& \qw 	& \\
      								& \qw & \qw 		& \qw 	& \qw 	& \qw 	& \qw 	& \qw 	& \targ	& \targ	& \qw 	& \qw 	& \qw	& \qw 	& \qw 	& \\
      								& \qw & \qw 		& \qw 	& \qw 	& \qw 	& \targ	& \targ	& \qw 	& \qw 	& \qw 	& \qw 	& \qw	& \qw 	& \qw 	& \\ 
      								& \qw & \qw	 	& \qw 	& \targ	& \targ	& \qw 	& \qw 	& \qw 	& \qw	& \qw 	& \qw 	& \qw	& \qw 	& \qw 	& \\      
      								& \qw & \qw 		& \targ 	& \qw 	& \qw 	& \qw	& \qw 	& \qw 	& \qw 	& \qw 	& \qw 	& \targ	& \qw 	& \qw 	& \\      
      \push{\vert 0 \rangle \hspace{0.1cm}} 	& \qw & \gate{H} 	& \qw 	& \qw 	& \qw 	& \qw 	& \qw 	& \qw 	& \qw 	& \qw 	& \ctrl{-5} 	& \ctrl{-1}	& \gate{H}	& \meter{} \\ 
      \push{\vert 0 \rangle \hspace{0.1cm}} 	& \qw & \gate{H} 	& \qw 	& \qw 	& \qw 	& \qw 	& \qw 	& \qw 	& \ctrl{-5} 	& \ctrl{-6} 	& \qw 	& \qw	& \gate{H}	& \meter{} \\
      \push{\vert 0 \rangle \hspace{0.1cm}} 	& \qw & \gate{H}	& \qw 	& \qw 	& \qw 	& \qw 	& \ctrl{-5} 	& \ctrl{-6}	& \qw 	& \qw 	& \qw 	& \qw	& \gate{H}	& \meter{} \\
      \push{\vert 0 \rangle \hspace{0.1cm}} 	& \qw & \gate{H}	& \qw 	& \qw 	& \ctrl{-5} 	& \ctrl{-6} 	& \qw 	& \qw 	& \qw 	& \qw 	& \qw 	& \qw	& \gate{H}	& \meter{} \\      
      \push{\vert 0 \rangle \hspace{0.1cm}} 	& \qw & \gate{H}	& \ctrl{-5} 	& \ctrl{-6} 	& \qw 	& \qw 	& \qw 	& \qw 	& \qw 	& \qw 	& \qw 	& \qw	& \gate{H}	& \meter{} \\
      \put(-0.4,342.5){\footnotesize{$1,j$}} 
      \put(-0.4,326.0){\footnotesize{$2,j$}} 
      \put(-0.4,309.5){\footnotesize{$3,j$}} 
      \put(-0.4,293.0){\footnotesize{$4,j$}} 
      \put(-0.4,276.5){\footnotesize{$5,j$}} 
}\]
\vspace{-15pt}
\caption{$Z$ syndrome extraction under $\mathcal{C}$($5$,$3$) for column $j$. To ensure fault tolerance it is necessary to repeat this circuit or to perform other redundant parity checks  \cite{Aliferis1}.}
\label{figure:zcorr5}
\end{figure}

\begin{figure}
\[ \Qcircuit @C=0.4em @R=0.36em @!R {                  
      								& \qw & \qw 		& \qw 	& \qw 	& \qw	& \ctrl{4} 	& \ctrl{3}	& \qw 	& \qw & \\
      								& \qw & \qw 		& \qw 	& \ctrl{4} 	& \ctrl{3}	& \qw	& \qw 	& \qw	& \qw & \\ 
      								& \qw & \qw 		& \ctrl{3} 	& \qw 	& \qw 	& \qw 	& \qw 	& \ctrl{1} 	& \qw & \\
      \push{\vert 0 \rangle \hspace{0.1cm}} 	& \qw & \qw 		& \qw 	& \qw 	& \qw 	& \qw	& \targ 	& \targ	& \meter{} \\
      \push{\vert 0 \rangle \hspace{0.1cm}} 	& \qw & \qw 		& \qw 	& \qw	& \targ 	& \targ	& \qw 	& \qw 	& \meter{} \\      
      \push{\vert 0 \rangle \hspace{0.1cm}} 	& \qw & \qw 		& \targ	& \targ	& \qw	& \qw 	& \qw 	& \qw 	& \meter{} \\
      \put(-0.4,211.5){\footnotesize{$i,1$}} 
      \put(-0.4,195.0){\footnotesize{$i,2$}} 
      \put(-0.4,178.5){\footnotesize{$i,3$}} 
}\]
\vspace{-15pt}
\caption{$X$ syndrome extraction under $\mathcal{C}$($5$,$3$) for row $i$ \cite{Aliferis1}.}
\label{figure:xcorr3}
\end{figure}
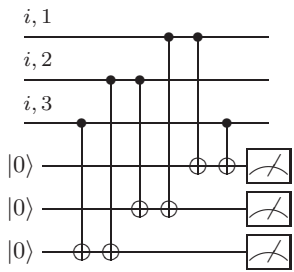

Here we consider the asymmetric subsystem code $\mathcal{C}$($5$,$3$), for which the circuits for $Z$ and $X$ syndrome extraction are shown in Figs.~\ref{figure:zcorr5} and \ref{figure:xcorr3} respectively \cite{Bacon1, Aliferis1}. With these circuits we are able to construct a memory extended rectangle and estimate the memory threshold \cite{Aliferis2}. An analysis of an asymmetric code requires that the usual threshold for arbitrary errors be separated into a $Z$ error threshold and an $X$ error threshold. We find that the $Z$ error threshold for a memory location under $\mathcal{C}$($5$,$3$) is approximately a factor of five higher than under $\mathcal{C}$($3$,$3$). This improvement is at the expense of the $X$ error threshold which is lowered by approximately the same factor. Additionally, if these threshold conditions are met, because it is a code that can correct two $Z$ errors, $\mathcal{C}$($5$,$3$) will afford a greater reduction in the failure rate than $\mathcal{C}$($3$,$3$), which can only correct a single $Z$ error. 

To enable universal fault tolerant quantum computation under $\mathcal{C}$($5$,$3$) we require circuits for a universal set of logical operations. As $\mathcal{C}$($5$,$3$) is a stabilizer CSS quantum code, logical $X$, $Z$, and {\sc cnot} are valid transversal operations. In addition to this set, logical $H$, $S$, and $T$ are required to form a universal set. However, due to the asymmetry in the stabilizer group, $\mathcal{S}$, of $\mathcal{C}$($5$,$3$), $H$ is not a valid transversal operation and so involves a more complex fault tolerant circuit. This problem is common to all asymmetric stabilizer CSS codes. Also $H$, $S$, and $T$ mix or transform $X$ and $Z$ errors, meaning that the asymmetry that $\mathcal{C}$($5$,$3$) is chosen to reflect is not preserved under these operations. We know that $H$ is a valid transversal operation under $\mathcal{C}$($3$,$3$) and we know that fault tolerant circuits for $S$ and $T$ are easily constructed if $H$ is already available \cite{Aliferis1}. Therefore, we propose to convert between $\mathcal{C}$($5$,$3$) and $\mathcal{C}$($3$,$3$) such that the better performing asymmetric code is used whenever possible and the symmetric code only when the circuit requires - that is, for the logical gates $H$, $S$, and $T$. We note that in some settings it may also be desirable to convert between symmetric codes to, for example, reallocate resources at different times during a large quantum computation. 

To convert between codes we are required to transform any valid logical state in one code to a valid logical state in the other and vice versa while preserving the information stored in the subsystem $\mathcal{H}_{\mathcal{L}}$.  The circuits that we present here satisfy these requirements for conversion between $\mathcal{C}$($3$,$3$) and $\mathcal{C}$($5$,$3$) and can be generalized to achieve conversion between larger subsystem codes.

\begin{figure}[b!]
\[ \Qcircuit @C=0.4em @R=0.36em @!R { 
									& \qw 	& \qw 	& \qw 	& \qw 	& \qw 	& \qw 	& \qw	\\
				 				         & \qw 	& \qw 	& \qw 	& \qw	& \targ	& \qw 	& \qw	\\
								         & \qw 	& \qw 	& \targ 	& \qw 	& \qw 	& \qw 	& \qw	\\
\push{\vert {0} \rangle \hspace{0.1cm}}		& \qw  	& \qw 	& \qw 	& \targ	& \qw 	& \qw 	& \qw	\\
\push{\vert {0} \rangle \hspace{0.1cm}}		& \qw  	& \targ 	& \qw 	& \qw	& \qw 	& \qw 	& \qw	\\
\push{\vert {0} \rangle \hspace{0.1cm}}		& \gate{H}	& \qw	& \qw	& \ctrl{-2}	& \ctrl{-4}	& \gate{H} & \meter{}	& & \\
\push{\vert {0} \rangle \hspace{0.1cm}}		& \gate{H}	& \ctrl{-2} 	& \ctrl{-4} 	& \qw	& \qw	& \gate{H} & \meter{}	& & \\ 
      \put(-0.4,244.5){\footnotesize{$1,j$}} 
      \put(-0.4,228.0){\footnotesize{$2,j$}} 
      \put(-0.4,211.5){\footnotesize{$3,j$}}
      \put(87,244.5){\footnotesize{$1,j$}} 
      \put(87,228.0){\footnotesize{$2,j$}} 
      \put(87,211.5){\footnotesize{$3,j$}} 
      \put(87,195.0){\footnotesize{$4,j$}} 
      \put(87,178.0){\footnotesize{$5,j$}}
      }\]     
\vspace{-18pt}
\caption{Conversion from $\mathcal{C}$($3$,$3$) to $\mathcal{C}$($5$,$3$) for column $j$.}
\label{figure:switch1}
\end{figure}
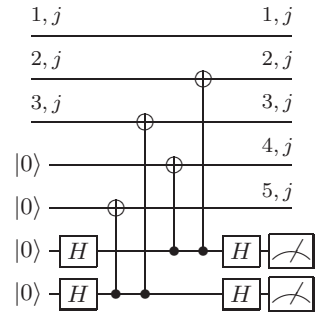

To convert from $\mathcal{C}$($3$,$3$) to $\mathcal{C}$($5$,$3$) we require stabilizing an additional two rows to the original grid, converting the stabilizer set, $\mathcal{S}$, from $\langle X_{i,*} X_{i+1,*} ; Z_{*,j}Z_{*,j+1} \ | \ i \in \Z_{2} ; \ j \in \Z_{2} \rangle$ to $\langle X_{i,*} X_{i+1,*} ; Z_{*,j}Z_{*,j+1} \ | \ i \in \Z_{4} ; j \in \Z_{2} \rangle$.  This can be done by initializing each of the two additional rows in the state $\vert 0^{\otimes3}\rangle$ and then measuring the parity of the operators $\langle X_{2,j} X_{4,j} ; X_{3,j}X_{5,j}\ | \ j \in \Z_{3} \rangle$ by the circuit in Fig.~\ref{figure:switch1}.  Note that each of the measured operators is supported on the new gauge group, $\mathcal{T}(5,3)$, and that the circuit in Fig.~\ref{figure:switch1} is actually a partial error correction protocol for $\mathcal{C}(5,3)$. Also, since the new rows are initialized in the state $\vert 0^{\otimes3}\rangle$ they are already stabilized with respect to $Z$ in both the stabilizer group and the gauge group. After measuring $\langle X_{2,j} X_{4,j}\rangle$ and $\langle X_{3,j}X_{5,j}\rangle$, the combined results are used to apply $Z$ corrections to qubits $(4,1)$ and $(5,1)$ respectively to ensure the state is a $+1$ eigenstate of the $\mathcal{C}$($5$,$3$) stabilizer set. A $Z$ error in row two or three of the initial $\mathcal{C}$($3$,$3$) logical state will be copied to row four or five of the final $\mathcal{C}$($5$,$3$) logical state respectively. Although this will result in two $Z$ errors, the process of converting from $\mathcal{C}$($3$,$3$) to $\mathcal{C}$($5$,$3$) is followed by correction under $\mathcal{C}$($5$,$3$) and so this is acceptable.

\begin{figure}
\begin{center}
\resizebox{45mm}{!}{\includegraphics{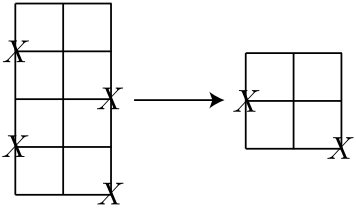}}
\end{center}
\vspace*{-10pt}
\caption{Converting from $\mathcal{C}$($5$,$3$) to $\mathcal{C}$($3$,$3$) requires eliminating this type of gauge state.  Here, the gauge part of the subsystem code is stabilized with respect to $X_{2,1}X_{4,1}$ and $X_{3,3}X_{5,3}$.  If the 6 qubits in the last two rows are measured, then 
residual $X$ errors remain at grid locations (2,1) and (3,3).  To protect against this type of error, the system is stabilized with respect to $Z_{4,1}Z_{4,2}$ and $Z_{5,2}Z_{5,3}$, this ensures that these $X$ gauge stabilizers are removed (since they do not commute with  $Z_{4,1}Z_{4,2}$ and $Z_{5,2}Z_{5,3}$) and errors are not induced during the conversion.}
\label{figure:set2}
\end{figure}
\begin{figure}[h!]
\begin{center}
\resizebox{49mm}{!}{\includegraphics{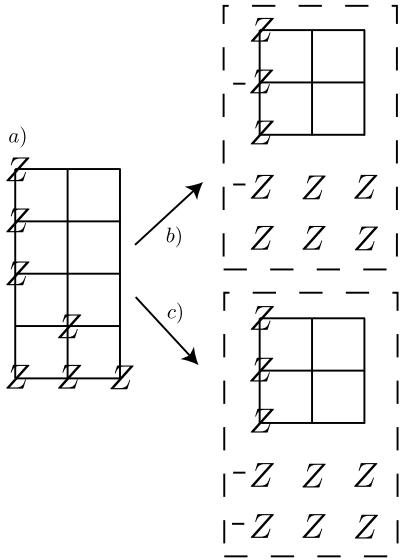}}
\end{center}
\vspace*{-10pt}
\caption{Converting from $\mathcal{C}$($5$,$3$) to $\mathcal{C}$($3$,$3$) can also enact a logical $Z$ operation.  Prior to the conversion, the system is stabilized with respect to $\langle Z_{i,1}Z_{i,2} ; Z_{i,2}Z_{i,3} \ | \ i \in \{4,5\} \rangle$.  Consider the case when the state is a $+1$ eigenstate of $Z_{*,1}$, (i.e. a $\vert+_L\rangle$ state), then Fig. a) represents the $Z$ operators for which the state is a $+1$ eigenstate.  In the absence of errors, after the conversion there are four possible measurement results of the six measured qubits.  Situation b) is where one row is measured $|111\rangle$ and the other $|000\rangle$.  In this case the converted state flips from a $+1$ eigenstate of $Z_{*,1}$ to a $-1$ eigenstate, therefore a logical $Z$ operation has been performed.  In situation c), both rows have been measured to be $|111\rangle$, since the factors of $-1$ cancel, the down converted state is still a $+1$ eigenstate of $Z_{*,1}$, therefore no logical operation has been performed.}
\label{figure:set3}
\end{figure}

Performing the conversion from $\mathcal{C}$($5$,$3$) to $\mathcal{C}$($3$,$3$) is slightly more complex.  The last two rows are disentangled by measuring each of the six qubits in the computational basis.  However, the specific gauge state of these rows must be fixed prior to conversion for two reasons:  First, if the gauge state of the system is fixed by $X_{i,j}X_{i',j} \in \mathcal{T}(5,3)$, where $i \in \{1,2,3\}$ and $i' \in \{4,5\}$, then measuring out the last two rows will leave one or more residual $X$ errors in the final $\mathcal{C}$($3$,$3$) logical state.  Secondly, the measurement result of the last two rows may induce a logical error if the gauge state of these rows is not known.  Prior to conversion, fixing the the last two rows to be $+1$ eigenstates of the operators $\langle Z_{i,1}Z_{i,2} ; Z_{i,2}Z_{i,3} \ | \ i \in \{4,5\} \rangle \in \mathcal{T}(5,3)$, using the circuit in Fig.~\ref{figure:xcorr3}, solves these two problems.  Fixing the gauge in this way ensures that the code block is not only stabilized by the $Z$ operators in $\mathcal{S}$ of $\mathcal{C}(5,3)$, but {\em also} by the $Z$ operators in $\mathcal{S}$ of $\mathcal{C}(3,3)$ - the classically controlled $X$ gates which are applied to the last two rows based on the parity result are also applied to the appropriate column in the first row to ensure the state is a $+1$ eigenstate of the $Z$ stabilizers of both $\mathcal{C}(5,3)$ and $\mathcal{C}(3,3)$.  As the last two rows are stabilized by $\langle Z_{i,1}Z_{i,2} ; Z_{i,2}Z_{i,3} \ | \ i \in \{4,5\} \rangle$, each measured row will have the same parity: $\vert 000\rangle$ or $\vert 111\rangle$.  As the logical $X$ operation is given by $Z_{*,1}$ for both codes, measuring one of the rows as $\vert 111\rangle$ is equivalent to applying a logical $Z$ gate to the final $\mathcal{C}$($3$,$3$) logical state. The measurement results of the last two rows can, therefore, enact the next logical $Z$ gate or this operation can be corrected. A majority vote of the measurements in each row ensures fault tolerance. Figs.~\ref{figure:set2} and~\ref{figure:set3} help to illustrate.  

\begin{figure}
\[ \Qcircuit @C=0.4em @R=0.36em @!R { 
										& \qw & \qw & / \qw & \qw & \qw & \qw & \qw 				& \qw 			& \gate{X} 		& \gate{X} 		& \gate{X^{\otimes3}}	& \qw & \qw & \qw & \qw & \qw & \qw & \qw & \qw & \qw & \qw \\	
										& \qw & \qw & / \qw & \qw & \qw & \qw & \qw 				& \qw 			& \qw \cwx		& \qw \cwx		& \qw \cwx	    		& \qw & \qw & \qw & \qw & \qw & \qw & \qw & \qw & \qw & \qw \\
										& \qw & \qw & / \qw & \qw & \qw & \qw & \qw 				& \qw 			& \qw \cwx		& \qw \cwx		& \qw \cwx			& \qw & \qw & \qw & \qw & \qw & \qw & \qw & \qw & \qw & \qw \\
										& \qw & \qw & / \qw & \qw & \qw  & \qw& \qw 				& \gate{\mathcal{P}} 	& \qw \cwx 		& \gate{X} \cwx		& \meter{} \cwx 			& \\
										& \qw & \qw & / \qw & \qw & \qw  & \qw& \gate{\mathcal{P}}	& \qw  			& \gate{X} \cwx 		& \qw \cwx 		& \meter{} \cwx[-1] 		& \\
 \push{\vert {0^{\otimes3}} \rangle \hspace{0.1cm}}	& \qw & \qw & / \qw & \qw & \qw & \qw & \qw 				& \gate{}\qwx[-2]	& \qw \cwx		& \meter{} \cwx[-1] 	& \\
 \push{\vert {0^{\otimes3}} \rangle \hspace{0.1cm}} 	& \qw & \qw & / \qw & \qw & \qw & \qw & \gate{} \qwx[-2]		& \qw 			& \meter{} \cwx[-1] 	& 			    	& \\ 
      \put(-0.4,247.5){\footnotesize{$i=1$}} 
      \put(-0.4,231.0){\footnotesize{$i=2$}} 
      \put(-0.4,214.5){\footnotesize{$i=3$}}
      \put(-0.4,198.0){\footnotesize{$i=4$}} 
      \put(-0.4,181.0){\footnotesize{$i=5$}}
      \put(154,247.5){\footnotesize{$i=1$}} 
      \put(154,231.0){\footnotesize{$i=2$}} 
      \put(154,214.5){\footnotesize{$i=3$}} 
      }\]     
\vspace{-18pt}
\caption{Conversion from $\mathcal{C}$($5$,$3$) to $\mathcal{C}$($3$,$3$). $\mathcal{P}$ is the gate sequence in Fig.~\ref{figure:xcorr3}.}
\label{figure:switch2}
\end{figure}
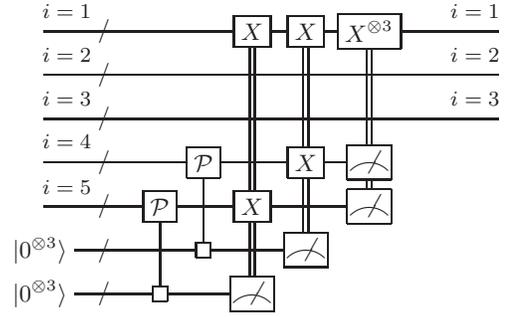

\begin{figure*}
\begin{center}
\resizebox{115mm}{!}{\includegraphics{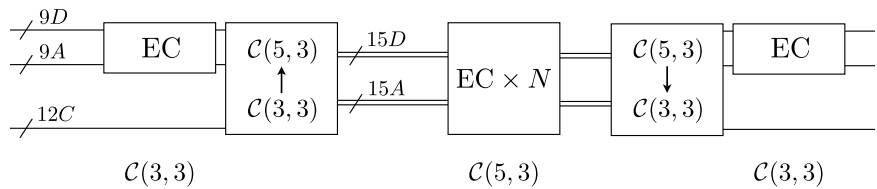}}
\end{center}
\vspace*{-15pt}
\caption{Using the combined symmetric asymmetric system for a period of logical memory. Blocks of qubits are labelled by the number of qubits in the block and by the role of the block at that part of the circuit, where $D$ indicates data, $A$ indicates ancilla, and $C$ indicates additional qubits that are required to convert to $\mathcal{C}$($5$,$3$). $N$ is the number of consecutive error correction cycles in the period. Estimates of the performance of this circuit relative to correction under $\mathcal{C}$($3$,$3$) are contained in Tab.~\ref{table:chains}.}
\label{figure:aec}
\end{figure*}

\begin{table*}
\begin{tabular}{c|c|c|c|c|c|c}
$\alpha=p_z/p_x$ & $N(\mathcal{R}=1)$ & $\mathcal{R}(N=10)$ & $\mathcal{R}(N=20)$ & $\mathcal{R}(N=50)$ & $\mathcal{R}(N=100)$ & $\mathcal{R}(N=1000)$ \\
$5$		& $14$	& $-$		& $1.0-1.2$	& $1.3-1.6	$	& $1.4-1.7$	& $1.5-2.0$ 	\\
$10$   	& $9$ 	& $1.0-1.2$ 	& $1.6-1.9$	& $2.3-3.2	$	& $2.7-4.1$	& $2.9-5.6$	\\
$100$  	& $7$ 	& $1.3-1.5$ 	& $2.2-2.9	$	& $4.0-7.2	$	& $5.4-14.0$	& $7.4-104.1$ 
\end{tabular} 
\caption{Estimates of the performance of the circuit in Fig.~\ref{figure:aec} relative to correction under $\mathcal{C}$($3$,$3$) for one level of error correction for various values of the ratio of the probabilities of $Z$ and $X$ errors, $\alpha$. The second column gives the number of successive logical memory locations, $N$, for greater than which it is beneficial to convert between $\mathcal{C}$($5$,$3$) and $\mathcal{C}$($3$,$3$). The remaining columns give the overall reduction in failure rate, $\mathcal{R}$, that is achieved for $N$ successive logical memory locations. Since $\mathcal{R}$ is dependent upon the absolute value of the error rate, $p_z$, we have specified its value in the range $0 < p_z \leq 10^{-4}$.}
\label{table:chains}
\end{table*}

To estimate how conversion between $\mathcal{C}$($5$,$3$) and $\mathcal{C}$($3$,$3$) affects the failure rate of a circuit we consider these conversion processes as logical operations and construct extended rectangles for each \cite{Aliferis2}. As any more than a single arbitrary error in each of these extended rectangles may cause circuit failure, we expect that the fidelity of the conversion operations will be lower than transversal operations under either $\mathcal{C}$($5$,$3$) or $\mathcal{C}$($3$,$3$). However, the penalty associated with converting between $\mathcal{C}$($3$,$3$) and $\mathcal{C}$($5$,$3$) will be offset by the benefit that is gained by computing under the better performing asymmetric code. The overall reduction in the failure rate that is achieved by using the combined symmetric asymmetric system will depend on the circuit and on the error asymmetry, and we note it will also not always be worth converting between codes. For a circuit that consists of $N$ successive logical memory locations we have estimated the reduction in failure rate, $\mathcal{R}$, that is achieved by converting between $\mathcal{C}$($3$,$3$) and $\mathcal{C}$($5$,$3$), as shown in Fig.~\ref{figure:aec}, for various values of the ratio of the probabilities of $Z$ and $X$ errors, $\alpha=p_z/p_x$. These estimates are contained within Table \ref{table:chains}.  

Note that in our analysis we have assumed a stochastic error model whereby every one- and two-qubit location will fail with equal probability and the error asymmetry, $\alpha$, is the same for all locations. In physical systems this will not generally be the case. A more realistic analysis of asymmetric error correction should account for the specific Hamiltonian from which gates are constructed and for the mixing of dephasing that occurs during the execution of non-diagonal gates such as CNOT.

The results presented in Table \ref{table:chains} are for one level of error correction only - for more levels of error correction any reduction in the failure rate after the first level will be compounded. Converting between $\mathcal{C}$($5$,$3$) and $\mathcal{C}$($3$,$3$) is suitable for an error asymmetry of around $\alpha=25$ with only one level of error correction, but for a larger error asymmetry it may be more beneficial to use two or more levels of asymmetric error correction or a more strongly asymmetric code at the first level. As fewer operations are transversal under an asymmetric code it may be preferable to use an asymmetric code at the first level only. Because some logical gates are able to be applied under $\mathcal{C}$($5$,$3$) without converting to $\mathcal{C}$($3$,$3$), it will be possible to design higher level error correction circuits that are amenable to the asymmetric quantum error correction at the first level. It is also possible to identify existing algorithmic circuits that contain extensive memory regions and so will benefit from asymmetric quantum error correction - for example the quantum Fourier transform and the quantum adder \cite{Meter1}. 

In conclusion, we have shown that using a combined symmetric asymmetric system allows benefit to be derived from asymmetric error correction without precluding universality. We acknowledge that further analysis of asymmetric errors is required before the benefit of this method is known for a specific physical setting, however the circuits presented in this paper may also be useful for other applications, such as converting ancilla state resources to error correction resources and vice versa.

\emph{Acknowledgments:} We acknowledge discussions with J. Cole, A. Fowler, A. Greentree, and C. Hill. SJD acknowledges the support of the David Hay Foundation. The work is funded by the Australian Research Council (ARC), the Australian Government, and by the US National Security Agency (NSA) and US Army Research Office (ARO) under Contract No. W911NF-04-1-0290.

\bibliographystyle{unsrt}

\end{document}